\documentclass[aps,prx,reprint,letterpaper,amsmath,amssymb,superscriptaddress,longbibliography]{revtex4-1}
\usepackage{bm}
\usepackage{graphicx}
\usepackage[colorlinks]{hyperref}
\usepackage[normalem]{ulem}

\begin{document}
\title{Quantum Interference Controls the Electron Spin Dynamics in $n$-GaAs}
\author{V.~V.~Belykh}
\email[]{vasilii.belykh@tu-dortmund.de}
\affiliation{Experimentelle Physik 2, Technische Universit\"{a}t Dortmund, D-44221 Dortmund, Germany}
\affiliation{P.N. Lebedev Physical Institute of the Russian Academy of Sciences, 119991 Moscow, Russia}
\author{A.~Yu.~Kuntsevich}
\affiliation{P.N. Lebedev Physical Institute of the Russian Academy of Sciences, 119991 Moscow, Russia}
\affiliation{National Research University Higher School of Economics, Moscow, 101000, Russia}
\author{M.~M.~Glazov}
\email[]{glazov@coherent.ioffe.ru}
\affiliation{Ioffe Institute, Russian Academy of Sciences, 194021 St. Petersburg, Russia}
\affiliation{Spin Optics Laboratory, St. Petersburg State University, 199034 St. Petersburg, Russia}
\author{K.~V.~Kavokin}
\affiliation{Ioffe Institute, Russian Academy of Sciences, 194021 St. Petersburg, Russia}
\affiliation{Spin Optics Laboratory, St. Petersburg State University, 199034 St. Petersburg, Russia}
\author{D.~R.~Yakovlev}
\affiliation{Experimentelle Physik 2, Technische Universit\"{a}t Dortmund, D-44221 Dortmund, Germany}
\affiliation{Ioffe Institute, Russian Academy of Sciences, 194021 St. Petersburg, Russia}
\author{M.~Bayer}
\affiliation{Experimentelle Physik 2, Technische Universit\"{a}t Dortmund, D-44221 Dortmund, Germany}
\affiliation{Ioffe Institute, Russian Academy of Sciences, 194021 St. Petersburg, Russia}

\begin{abstract}
Manifestations of quantum interference effects in macroscopic objects are rare. \emph{Weak localization} is one of the few examples of such effects showing up in the electron transport through solid state. Here we show that weak localization becomes prominent also in optical spectroscopy via detection of the electron spin dynamics. In particular, we find that weak localization controls the free electron spin relaxation in semiconductors at low temperatures and weak magnetic fields by slowing it down by almost a factor of two in $n$-doped GaAs in the metallic phase. The weak localization effect on the spin relaxation is suppressed by moderate magnetic fields of about 1~T, which destroy the interference of electron trajectories, and by increasing the temperature. The weak localization suppression causes an anomalous decrease of the longitudinal electron spin relaxation time $T_1$ with magnetic field, in stark contrast with well-known magnetic field induced increase in $T_1$. This is consistent with transport measurements which show the same variation of resistivity with magnetic field. Our discovery opens a vast playground to explore quantum magneto-transport effects optically in the spin dynamics.
\\
\\
\doi{10.1103/PhysRevX.8.031021}
\end{abstract}

\maketitle
\section{Introduction}
The design of future spintronic and opto-spintronic devices requires a detailed understanding of the correlation between the electron conductivity and spin relaxation in prospective material systems, such as semiconductors. The electron spin relaxation in semiconductors depends strongly on whether electrons are itinerant or localized~\cite{Dzhioev2002,Belykh2017}. Across the metal-to-insulator transition (MIT) the spin relaxation changes as dramatically as does the conductivity~\cite{Belykh2017,Lonnemann2017}. Indeed, in the insulating phase both conductivity and spin relaxation critically depend on the overlap of the wavefunctions of donor-bound electrons at low temperatures and on the number of delocalized electrons at higher temperatures. In the metallic phase, in semiconductors without an inversion center with GaAs as prototype system, the spin relaxation is governed by spin-orbit coupling (Dyakonov-Perel mechanism) \cite{Dyakonov1972} and, similarly to the conductivity, the spin relaxation becomes suppressed by electron scattering events. The spin relaxation rate is closely related to the electron diffusion coefficient~\cite{Kavokin08,Dyakonov86,Dyakonov2017}, so that charge transport phenomena are generally expected to manifest also in spin relaxation processes \cite{Sih2004,Fukuoka2008,Larionov2015,Larionov2017}. The situation becomes particularly involved in the vicinity of the MIT where quantum effects become important~\cite{Shklovskii06,Lyubinskiy2004}.

While the mechanisms of electron spin relaxation in semiconductors were largely clarified in theory back in the 1970s~\cite{Pikus1984}, for a long time experiments could access the electron spin dynamics only via the Hanle effect near zero magnetic field. Since the 1990s more advanced techniques have become available such as pump-probe methods analyzing the Kerr/Faraday rotation \cite{Baumberg1994,Zheludev1994} or polarization-resolved photoluminescence \cite{Colton2004,Colton2007,Fu2006,Linpeng2016}, and elaborated methods like resonant spin amplification \cite{Kikkawa1998,Yugova2012}, spin noise spectroscopy \cite{Oestreich2005,Crooker2009,Romer2010} and spin inertia reorientation \cite{Heisterkamp2015}. Each of these tools has limitations related to the achievable time resolution, the addressable time range or the applicable magnetic field. So far, access to the relation between the electron diffusion and spin relaxation in the vicinity of the MIT was hindered by experimental limitations. Only recently the pump-probe technique was extended to facilitate direct measurements of arbitrarily long spin dynamics with picosecond time resolution across a wide range of magnetic fields \cite{Belykh2016}.

On the other hand, the transport properties of semiconductors that directly provide information about electron diffusion, are rather easily accessible in experiment. In weak magnetic fields the low temperatures magnetoresistance is negative due to the {\it weak localization effect}: the magnetic field destroys the phase coherence of interfering paths and increases the electron diffusion coefficient~\cite{Altshuler1985,Fritzsche1955,Woods1964,Halbo1968,Benzaquen1988,Kawabata1980,Kawabata1980a,Capoen1993}. The spin-orbit interaction has a pronounced impact on the low-field magnetoresistance leading to positive magnetoresistance, i.e., antilocalization, if the spin coherence of electrons is lost faster than their phase~\cite{Altshuler1985}. Although weak localization/antilocalization is expected to emerge in the spin dynamics~\cite{Lyubinskiy2004,Lyubinskiy2005}, it has not been identified in experiments so far.

In this paper we demonstrate that weak localization significantly slows down the itinerant electron spin relaxation in the Dyakonov-Perel' mechanism. Using the extended pump-probe Faraday rotation technique we study the longitudinal electron spin relaxation time $T_1$ as a function of external magnetic field in $n$-doped metallic bulk GaAs. While the classical theory \cite{Ivchenko1973} predicts an increase of $T_1$ with increasing field mainly due to the cyclotron motion of the free carriers, we observe an anomalous decrease of $T_1$ in moderate fields $B\lesssim 1$~T. From transport measurements done on the same samples we observe that the negative magnetoresistance is correlated with the anomalous magnetic field dependence of $T_1$.
We develop a theoretical model of the weak localization effect in the spin relaxation of bulk semiconductors and find very good agreement between the calculations and experimental data. Our results establish a strict relation between the electron diffusion and spin relaxation in metallic systems in the vicinity of the MIT. Thereby all-optical access to weak localization is provided and a tool to probe \emph{locally} electron transport phenomena is developed.

\section{Experimental details}
The results are obtained on Si-doped GaAs samples with electron concentrations of $n_\text{e} = 5.5 \times 10^{14}$~cm$^{-3}$ (2-$\mu$m-thick layer grown by the molecular-beam epitaxy), $3.7 \times 10^{16}$~cm$^{-3}$ and $7.1 \times 10^{16}$~cm$^{-3}$ (140 and 170-$\mu$m-thick bulk wafers, respectively).

For optical measurements the samples are placed in the variable temperature insert of a split-coil magnetocryostat ($T =2-25$~K). Magnetic fields up to 6~T are applied parallel to the light propagation direction that is parallel to the sample growth axis (Faraday geometry). The extended pump-probe Kerr/Faraday rotation technique is used to study the electron spin dynamics. It is a modification of the standard pump-probe Kerr/Faraday rotation technique, where circularly-polarized pump pulses generate carrier spin polarization, which is then probed by the Kerr~(Faraday) rotation of linearly-polarized probe pulses after reflection~(transmission) from~(through) the sample. Implementation of pulse picking for both pump and probe beams in combination with a mechanical delay line allows us to scan microsecond time ranges with picosecond time resolution. Details of the technique are given in Ref.~\cite{Belykh2016}.

Here, a Ti:Sapphire laser emits a train of 2~ps pulses with a repetition rate of 76~MHz (repetition period $T_\text{R}=13.1$~ns). The pump protocol uses single pulse per excitation period. The separation between these pulses is $80 T_\text{R}$, $160 T_\text{R}$ or $320 T_\text{R}$ in order to clearly exceed the characteristic time of spin polarization decay. The sample with donor concentration $n_\text{e}$ of $5.5 \times 10^{14}$~cm$^{-3}$ is studied in reflection geometry (Kerr rotation) with the laser wavelength set to 819~nm, close to the donor-bound exciton resonance. The samples with $n_\text{e} = 3.7 \times 10^{16}$~cm$^{-3}$ and $7.1 \times 10^{16}$~cm$^{-3}$ are studied in transmission geometry (Faraday rotation) with the laser wavelength set to 829~nm.

Magnetoresistance measurements were performed using a standard 4-terminal technique with a lock-in amplifier. The measurement current (36 Hz, 100 $\mu$A) was checked not to overheat the sample at the lowest temperature. Ohmic contacts (with an almost $T$-independent resistance of about 100 Ohm) were obtained by annealing of indium drops on top of the preliminary scratched wafer (10 minutes at 400$^o$C in vacuum). A PPMS-9 cryostat and Cryogenics CFMS-16 system were used to set the temperature (2-40~K) and magnetic field (up to 6~T). The magnetic field perpendicular to the sample surface and current direction was swept from positive to negative values with subsequent symmetrization of the data to compensate inevitable contact misalignment.

\section{Results and discussion}
\begin{figure}
\includegraphics[width=0.8\columnwidth]{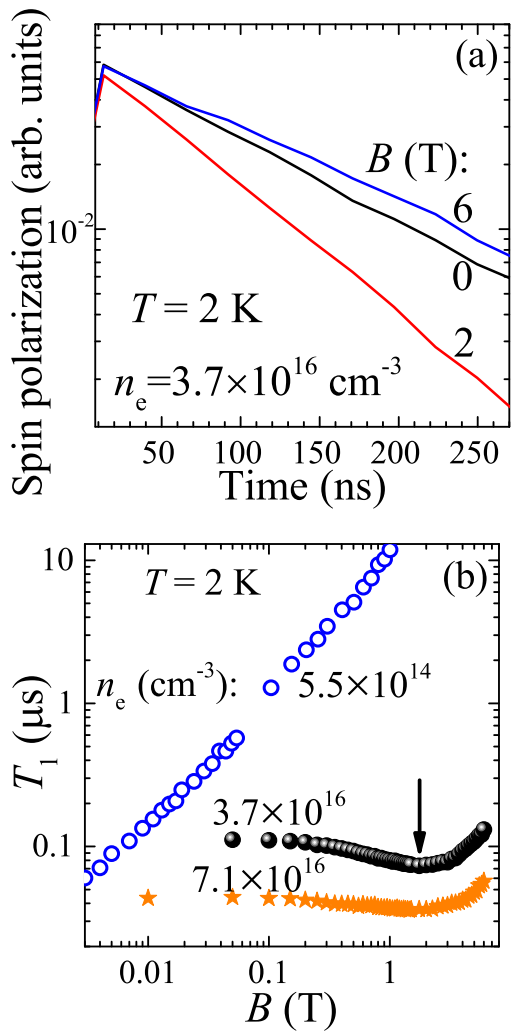}
\caption{Longitudinal spin relaxation. (a) Dynamics of the electron spin polarization (measured as Faraday rotation signal) at different magnetic fields for the $n$-GaAs sample with $n_\text{e} = 3.7 \times 10^{16}$~cm$^{-3}$. (b) Magnetic field dependence of the longitudinal relaxation time $T_1$ for samples with different donor concentrations. The arrow indicates the minimum in the $T_1(B)$ dependence. (a),(b) Temperature $T = 2$~K.}
\label{fig:Kin}
\end{figure}

\emph{Experiment}. The circularly polarized pump laser pulse creates spin polarization along the magnetic field (Faraday geometry $\bm B \parallel \bm z \parallel [001]$) which can be detected by the delayed probe laser pulse via Faraday rotation of its linear polarization. Figure~\ref{fig:Kin}(a) shows the dynamics of the spin polarization for exemplary values of the external magnetic field $B=0$, 2 and 6~T for the metallic sample with electron concentration $n_\text{e} = 3.7 \times 10^{16}$~cm$^{-3}$, which is somewhat above the MIT threshold, $n_\text{e}^{\rm MIT} \approx (1-2) \times 10^{16}$~cm$^{-3}$. The signal decays monoexponentially with the longitudinal spin relaxation time $T_1$. It is seen from Fig.~\ref{fig:Kin}(a) that as the magnetic field grows, $T_1$ first decreases, reaches a minimum and then increases. The non-monotonic dependence of $T_1(B)$ with a minimum at about $1.5$~T is further substantiated in Fig.~\ref{fig:Kin}(b) by the solid spheres. The minimum in the $T_1(B)$ dependence becomes less pronounced for the sample with even higher carrier concentration, while for the samples with lower donor concentrations, below MIT, $T_1$ monotonically increases with increasing $B$ [the open circles in Fig.~\ref{fig:Kin}(b)].

\begin{figure}
\includegraphics[width=1\columnwidth]{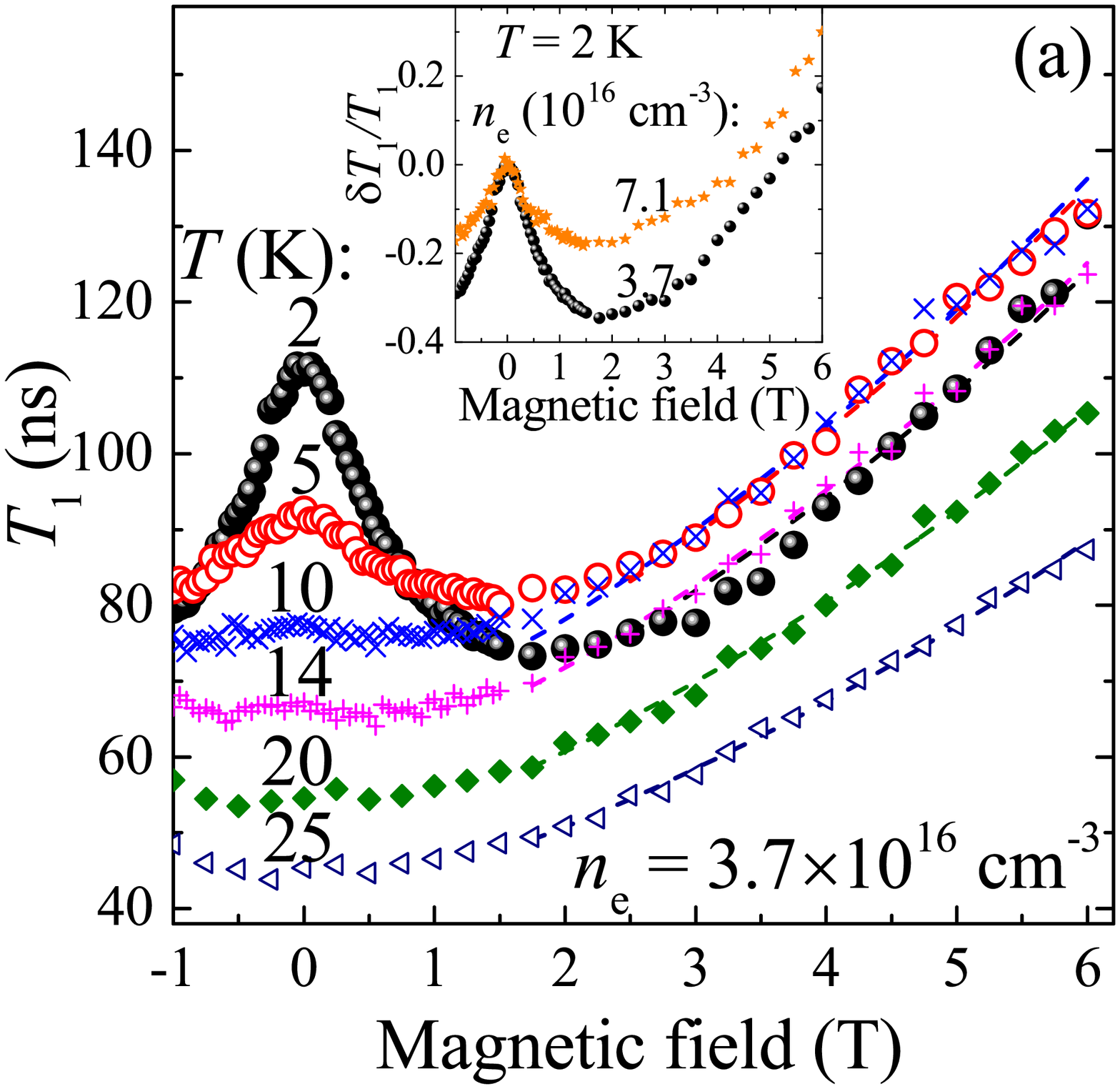}
\includegraphics[width=1\columnwidth]{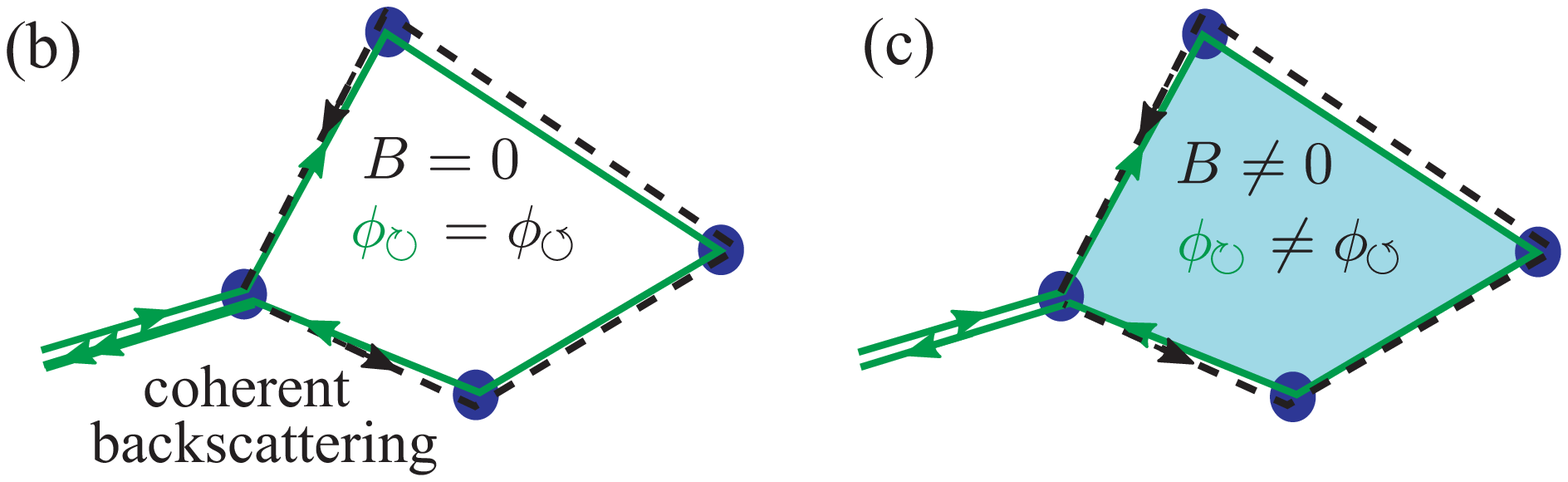}
\caption{Effect of weak localization on longitudinal spin relaxation time $T_1$. (a) Magnetic field dependence of $T_1$ at different temperatures. $n_\text{e} = 3.7 \times 10^{16}$~cm$^{-3}$. The dashed lines show fits to the experimental data with Eq.~\eqref{eq:Ivch}. Inset shows relative variation of $T_1$ with magnetic field for metallic samples with different electron concentrations. (b) Scheme of constructive interference of clockwise and counter-clockwise electrons paths, starting at the same impurity and related by the time reversal symmetry. The interference gives rise to the weak localization effect by increasing the backscattering efficiency. (c) The interference between the same paths as in Fig.~\ref{fig:T1}(b) is destroyed by the magnetic field due to the extra phase acquired by the electron traveling clockwise and counter-clockwise.}
\label{fig:T1}
\end{figure}

To investigate the anomalous $T_1(B)$ dependence further we perform measurements at different temperatures with the results summarized in Fig.~\ref{fig:T1}(a). Interestingly, the minimum in the $T_1(B)$ dependence at increased $B$ (or, alternatively, peak at $B=0$) is observed only at low temperatures $T \lesssim 14$~K. Furthermore, with increasing temperature the minimum becomes less pronounced due to the decrease of the zero-field $T_1$ value. The decrease of $T_1$ with magnetic field or temperature increase are unexpected in view of existing theories of free-electron spin relaxation in semiconductors~\cite{Dyakonov1972,Ivchenko1973,Pikus1984,Dyakonov2017}. This calls for a detailed modeling of the spin relaxation process which is presented below.

\emph{Model.} In GaAs-like semiconductors, being in the metallic phase, the spin relaxation is controlled by the Dyakonov-Perel mechanism~\cite{Dyakonov1972,Pikus1984}: the electron spin precesses around the effective, spin-orbit coupling-induced magnetic field and the spin precession is randomized by scattering events. The spin dynamics is described in the framework of a kinetic equation for the spin distribution function $\bm s_k$~\cite{Dyakonov1972,Ivchenko1973,Glazov02,Glazov04}
\begin{equation}
\label{eq:ke}
\frac{\partial \bm s_{\bm k}}{\partial t} + {\Lambda}_{\bm k}\{ \bm s_{\bm k}\} + \bm s_{\bm k} \times \bm \Omega_{\bm k} = Q\{\bm s_{\bm k}\}.
\end{equation}
Each term in Eq.~\eqref{eq:ke} has a transparent physical meaning. The operator ${\Lambda}_{\bm k}= -\bm \omega_c[\bm k\times \partial/\partial \bm k]$ describes the electron cyclotron motion in the external magnetic field, where $\bm \omega_c = e\bm B/mc$ is the cyclotron frequency, $m$ is the electron effective mass and $e$ is the electron charge (the Zeeman splitting is neglected). The term $\bm s_{\bm k} \times \bm \Omega_{\bm k}$ describes the precession of the electron spin around the effective magnetic field arising due to the spin-orbit interaction in a system with bulk inversion asymmetry. The corresponding precession frequency $\bm\Omega_{\bm k}$ is cubic on the electron wavevector $\bm k$. In the last term in Eq.~\eqref{eq:ke}, $Q\{\bm s_{\bm k}\}$ is the collision integral, i.e., the operator describing the redistribution of electrons between different states in $\bm k$-space. It takes into account the electron scattering and can be generally presented as
\begin{equation}
\label{Qk}
Q\{\bm s_{\bm k}\} = \sum_{\bm k'} \left(W_{\bm k\bm k'} \bm s_{\bm k'} - W_{\bm k'\bm k} \bm s_{\bm k}\right),
\end{equation}
describing the balance between the processes where an electron leaves the state with wavevector $\bm k'$ and is promoted to the state with wavevector $\bm k$ with the rate $W_{\bm k\bm k'}$ and vice versa, accordingly. For the elastic scattering by the central potential of ionized donors relevant for the studied system, $W_{\bm k\bm k'} = W_{\bm k'\bm k}$, and relaxation of different angular harmonics $Y_{lm}(\vartheta_{\bm k},\varphi_{\bm k})$ of the distribution function ($\vartheta$ and $\varphi$ are the angles of the wavevector) occurs independently~\cite{Dyakonov1972}. Thus, for the spin distribution $\bm s_{\bm k} = \delta \bm s_k Y_{lm}(\vartheta_{\bm k},\varphi_{\bm k})$ the collision integral $Q\{\delta \bm s_{k} Y_{lm}(\vartheta_{\bm k},\varphi_{\bm k})\} = -\tau_l^{-1} \delta \bm s_{k} Y_{lm}(\vartheta_{\bm k},\varphi_{\bm k})$ and it is described by a set of relaxation times $\tau_{l}$ ($l=1,2,3,\ldots$):
\begin{equation}
\label{tau:l}
\frac{1}{\tau_l} = \sum_{\bm k'} W_{\bm k'\bm k} [1-P_l(\cos{\vartheta_{\bm k'}})],
\end{equation}
responsible for the relaxation of different angular harmonics of the distribution function; $P_l(x)$ is the corresponding Legendre polynomial. Note, that $\tau_1 = \tau_p$ describes the momentum relaxation of electrons. In a classical approach, these relaxation times are independent of the magnetic field.

The electron scattering slows down the spin relaxation due to randomization of the spin precession around the spin-orbit magnetic field: between the scattering acts the electron spin rotates by a small angle $\sim \Omega_{\bm k} \tau$ ($\tau$ is the characteristic relaxation time), while the scattering processes changes the wavevector $\bm k$ and, correspondingly, the spin precession frequency $\bm\Omega_{\bm k}$ reducing the cumulative spin rotation angle. It follows from the solution of Eq.~\eqref{eq:ke} that the longitudinal spin relaxation time for degenerate electrons in bulk GaAs at $B = 0$ takes the form~\cite{Dyakonov1972}
\begin{equation}
\label{eq:T0}
T_1(0) = \frac{105}{32\alpha^2} \frac{\hbar^2 E_\text{g}}{E_\text{F}^3 \tau_3},
\end{equation}
where $E_\text{F}$ is the electron Fermi energy, $E_\text{g}=1.52$~eV is the band gap energy, $\alpha\approx 0.063$ is the dimensionless Dresselhaus constant for GaAs recalculated from data in Refs.~\cite{Jusserand1995,Richards1999}, and $\tau_3$ is the relaxation time of third angular harmonics of the electron  distribution over momentum given by Eq.~\eqref{tau:l}.

A similar suppression of the spin relaxation takes place due to the cyclotron motion of the electron in external magnetic field accounted for by the operator ${\Lambda}_{\bm k}\{ \bm s_{\bm k}\}$ in Eq.~\eqref{eq:ke}. Indeed, the field induces a rotation of the electron velocity and the wavevector $\bm k$, thus, resulting in a rotation of the effective magnetic field $\propto \bm\Omega_{\bm k}$. In this way, the magnetic field acts as an extra scattering source and slows down the spin relaxation~\cite{Ivchenko1973,Marushchak1983}.
The magnetic field dependence of $T_1$ was calculated in Ref.~\cite{Ivchenko1973}:
\begin{equation}
\label{eq:Ivch}
\frac{T_1(B)}{T_1(0)} = \frac{[1+(\omega_c\tau_3)^2] [1+9 (\omega_c\tau_3)^2]}{1+6 (\omega_c\tau_3)^2} \approx 1+4\omega_c^2\tau_3^2.
\end{equation}
The last approximate equality in Eq.~\eqref{eq:Ivch} holds for $\omega_c \tau_3 \ll 1$. Equation \eqref{eq:Ivch} clearly demonstrates an increase in the spin relaxation time $T_1$ with growing magnetic field. This expression with the temperature-independent $\tau_3 \approx 40$~fs describes the experimental data at $B\gtrsim 2$~T [the dashed lines in Fig.~\ref{fig:T1}(a)]. From the value of $T_1$ extrapolated to $B=0$, we obtain after Eq.~\eqref{eq:T0} almost the same $\tau_3$ as the value obtained above from the $B$-dependence.

The classical theory of Dyakonov-Perel spin relaxation mechanism, expressed by Eqs.~\eqref{eq:T0} and \eqref{eq:Ivch}, as well as additional possible mechanisms of spin relaxation due to the $g$-factor spread~\cite{Bronold2002} cannot, however, explain the sizable decrease of the spin relaxation time $T_1$ in rather weak magnetic fields $B \lesssim 1$~T and at low temperatures $T\lesssim 14$~K. Clearly, other effects, not accounted for by the approach in Refs.~\cite{Dyakonov1972,Ivchenko1973,Pikus1984,Dyakonov2017,Glazov02,Glazov04} must play an important role in our experiment. In fact, in the derivation of Eqs.~\eqref{eq:T0} and \eqref{eq:Ivch} the electron dynamics is assumed to be classical, i.e., the inequality $E_\text{F} \tau_p/\hbar \gg 1$ is assumed to hold. For relatively low electron densities, $E_\text{F}\tau_p/\hbar$ just slightly exceeds unity and quantum effects start to play a role. In particular, for an electron traveling through a disordered medium the interference between classical trajectories, as schematically depicted in Fig.~\ref{fig:T1}(b) becomes important. For electron waves traveling clockwise and counter-clockwise through the same configuration of impurities, the phases acquired on these two paths, $\phi_{\circlearrowright} = \phi_{\circlearrowleft} = \oint \bm k d\bm l$, are the same. As a result, the two paths shown by the solid and dashed lines interfere constructively, leading to coherent backscattering. In effect, the scattering efficiency by the impurities increases ($\tau_p$ decreases) and the electron propagation slows down. This is the \emph{weak localization} effect signifying the onset of the MIT with decreasing electron density. Importantly, a magnetic field destroys the constructive interference owing to the extra phase proportional to the field flux through the trajectory acquired by the diffusing electron, see Fig.~\ref{fig:T1}(c). Indeed, for clockwise and counter-clockwise propagation the field-induced phases are opposite, hence, the magnetic field suppresses the weak localization~\cite{Kawabata1980,Kawabata1980a,Capoen1993,Gorkov1979,Altshuler1985}.

In order to account for the interference effect we follow the semiclassical approach where, as illustrated in Fig.~\ref{fig:T1}(b),(c), the quantum effects are accounted for by renormalization of the cross-section for electron scattering by the impurity~\cite{Dmitriev1997,Lyubinskiy2004,Lyubinskiy2005}. The momentum relaxation time $\tau_p$ acquires a correction $\delta \tau_p$ of the form
\begin{equation}
\label{eq:transport}
\frac{\delta \tau_p}{\tau_p} = - \frac{2m D}{\pi \hbar n_\text{e}} \sum_{s,s'=\pm 1/2} \mathcal C_{ss's's}(\bm r=0),
\end{equation}
where $D = v_\text{F}^{2} \tau_p/3$ is the electron diffusion coefficient, and $v_\text{F}$ is the Fermi velocity. In Eq.~\eqref{eq:transport}, $\mathcal C_{s_1s_2s_3s_4}(\bm r=0)$ is the Cooperon matrix describing the electron interference along the closed loops, which is calculated via a standard diagram technique~\cite{Altshuler1985}.
Similarly, the interference effects modify the relaxation time of the third angular harmonics of the spin distribution function, $\tau_3$, which defines the spin relaxation time [Eq.~\eqref{eq:T0}], as:
\begin{equation}
\label{spin:harm}
\frac{\delta \tau_3}{\tau_3}
= - \frac{2 m D}{\pi \hbar n_\text{e}} \sum_{s,s'=\pm 1/2} \mathcal C_{ss's's}(\bm r=0) (2\delta_{ss'}-1).
\end{equation}
The factor $(2\delta_{ss'}-1)$ is due to the spin vortices in the corresponding diagrams~\cite{Lyubinskiy2004,Lyubinskiy2005}.

The Cooperon matrix, i.e., the sum of maximally crossed diagrams, describes the spin-dependent probability $P_{\rm ret}$ of an electron to return to the initial point after an arbitrary number of collisions conserving its phase, that is the probability to pass through a loop in the real space [Fig.~\ref{fig:T1}(b)]. Qualitatively, the interference of electron waves propagating clockwise and counterclockwise on the loops, Fig.~\ref{fig:T1}(b), gives rise to the coherent backscattering effect and modifies the rate of the scattering by an impurity $W_{\bm k\bm k'}$. It gives rise to a sharp peak in $W_{\bm k\bm k'}$ at $\bm k' \approx - \bm k$, i.e., for backscattering~\cite{Dmitriev1997}. The interference induced contribution $\delta W_{\bm k\bm k'} =  W_{\bm k\bm k'} - W^{cl}_{\bm k\bm k'}\propto P_{\rm ret}$, where $W^{cl}_{\bm k\bm k'}$ is the classical value found without interference effects, is proportional to the return probability. It gives rise to the corrections $\delta \tau_l$  to the relaxation times $\tau_l$ in Eq.~\eqref{tau:l}. Both $\delta W_{\bm k\bm k'}$ and $\delta \tau_l$ are determined by the interference of the trajectories in Fig.~\ref{fig:T1}(b). The magnetic field destroys the interference and suppresses $\delta\tau_l$ affecting the electron transport and spin dynamics.

We introduce the phase relaxation time $\tau_\phi$ associated with inelastic electron-electron or electron-phonon scattering processes, and consider hereafter the diffusive regime where $\tau_\phi \gg \tau_p,\tau_3$ and the magnetic length $l_B = \sqrt{\hbar c/(|e| B)}$ exceeds by far the mean free path, $l_B \gg v_{\rm F}\tau_p$. Moreover, we impose the condition of rather weak spin-orbit interaction, $T_1(0) \gg \tau_\phi$, meaning that the electron spin is conserved during passage through the closed loops in which the interference takes place. As a result, we have
\begin{multline}
\label{tau:s:b}
 \frac{\delta T_1(B)}{T_1(0)} = \frac{\delta \rho(B)}{\rho(0)}
= -{\frac{m}{2\pi^2\hbar n_\text{e} \tau_p} \sqrt{\frac{|e|B}{\hbar c}}} F_3\left(\frac{B_\phi}{4B}\right).
\end{multline}
Here $B_\phi = \hbar c/(|e| l_\phi^2)$, $l_\phi = \sqrt{D\tau_\phi}$ is the phase relaxation length, and the function $F_3(x)$ is defined as~\cite{Kawabata1980,Kawabata1980a,Capoen1993}
\begin{equation}
F_3(x) = \sum_{n=0}^\infty \left[ 2(\sqrt{n+1+x}-\sqrt{n+x}) - \frac{1}{\sqrt{n+1/2+x}}\right].\nonumber
\end{equation}
Note that for $x\ll 1$ $F_3(x)\approx0.605$ and for $x\gg 1$, $F_3(x) \approx 1/(48x^{3/2})$.

\emph{Discussion and comparison of electron spin dynamics and transport}

\begin{figure}
\includegraphics[width=1\columnwidth]{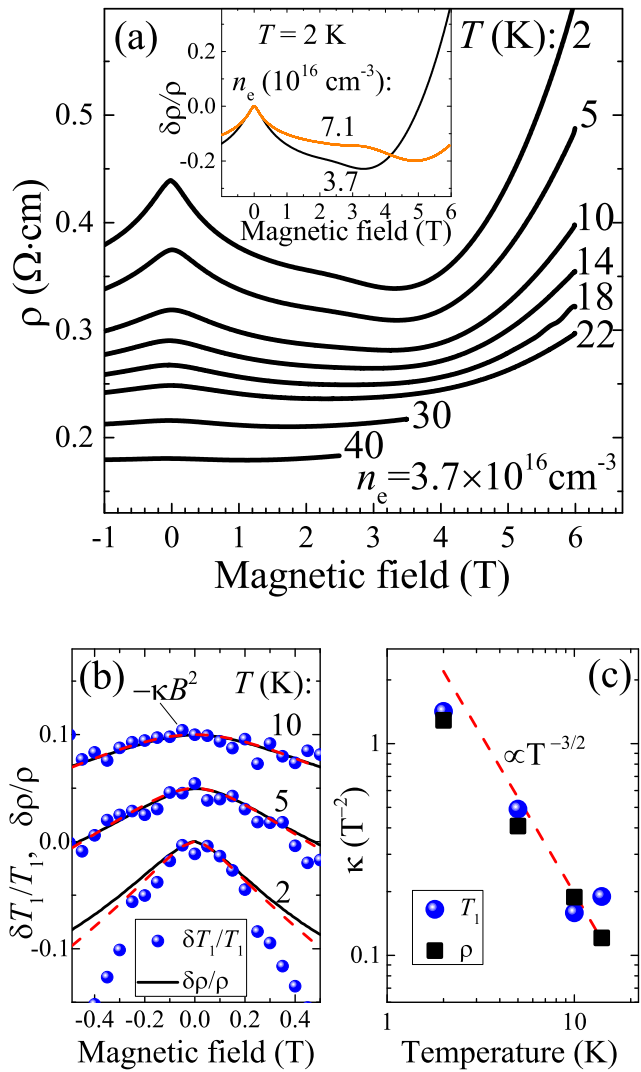}
\caption{Evidence of weak localization in resistivity measurements. (a) Magnetic field dependence of the resistivity $\rho$ at different temperatures. Inset shows relative variation of $\rho$ with magnetic field for metallic samples with different electron concentrations. (b) Relative variation of $T_1$ (the symbols) and $\rho$ (the solid lines) with magnetic field at different temperatures. The red dashed lines show fits to both $\delta T_1 / T_1$ and $\delta \rho / \rho$ with Eq.~\eqref{tau:s:b}. The curves are vertically shifted for clarity. (c) Curvatures of the magnetic field dependencies of $\delta T_1 / T_1$ (the spheres) and $\delta \rho / \rho$ (the squares), $\kappa$ in Eq.~\eqref{eq:LowB}, as a function of temperature. The red dashed line shows a $T^{-3/2}$ dependence. (a)--(c) $n_\text{e} = 3.7 \times 10^{16}$~cm$^{-3}$.}
\label{fig:rho}
\end{figure}

To independently experimentally confirm the presence of weak localization and estimate its magnitude in the considered system, we have also measured the {\it magnetoresitance} on the same samples [see Fig.~\ref{fig:rho}(a)]. The low-field negative magnetoresistance is clearly seen, in agreement with previous works it arises from the weak localization effect~\cite{Fritzsche1955,Woods1964,Halbo1968,Benzaquen1988,Kawabata1980,Kawabata1980a,Capoen1993,Emelyanenko1982}. At high fields positive magnetoresistance is observed, presumably due to the field-induced compression of electron wave functions on donors and also possibly due to the onset of Shubnikov-de Haas oscillations. The observed behavior is qualitatively similar to that for $T_1(B)$ [Fig.~\ref{fig:T1}(a)] and, in particular, the scale of magnetic field, destroying the weak localization, is the same. Further, the negative magnetoresistance persists in the same range of temperatures as the decrease of $T_1$ with $B$.

Furthermore, according to Eq.~\eqref{tau:s:b}, the relative change of $T_1$ and $\rho$ due to the weak localization should be the same. Figure~\ref{fig:rho}(b) shows these relative variations of $T_1$ (the spheres) and $\rho$ (the solid lines), $\delta T_1/T_1 \equiv T_1(B)/T_1(0) - 1$ and $\delta \rho/\rho \equiv \rho(B)/\rho(0) - 1$, with magnetic field, respectively. Equation~\eqref{tau:s:b} is, strictly speaking, valid if the quantum corrections are small, i.e. for $\delta T_1/T_1, \delta \rho/\rho \ll 1$ which is not the case in our sample right above the MIT. Nevertheless, the measured magnetic field dependences of $\delta T_1/T_1$ and $\delta \rho/\rho$ are in remarkable agreement in weak magnetic fields. The analysis of the asymptotic form of Eq.~\eqref{tau:s:b} shows that in weak fields $B \ll B_\phi$
\begin{equation}
\label{eq:LowB}
 \frac{\delta T_1(B)}{T_1(0)} = \frac{\delta \rho(B)}{\rho(0)} \approx -\kappa B^2,
\end{equation}
with the prefactor $\kappa \approx 0.048 (e/m c)^2 \sqrt{\tau_p\tau_\phi^{3}}$. In the studied temperature range $\tau_p$ is constant, as found above, and $\tau_\phi = A/T$, where $A$ is a constant, in accordance with Refs.~\cite{Capoen1993,Isawa1984}. Thus, $\kappa \propto T^{-3/2}$. The values of curvature $\kappa$ corresponding to $T_1$ and $\rho$ extracted from the fit are shown in \ref{fig:rho}(c). They are in very good agreement and follow a $T^{-3/2}$ dependence as shown by the red dashed line.

The dashed lines in Fig.~\ref{fig:rho}(b) show fits to the experiment by Eq.~\eqref{tau:s:b} using a reasonable set of parameters, namely $\tau_p = 55$~fs (temperature independent) and $\tau_\phi =A/T$ with $A = 19$~ps$\cdot$K. Such  inverse temperature dependence of the phase relaxation time was observed for a similar GaAs system~\cite{Capoen1993} with $n_\text{e} = 2.9 \times 10^{16}$~cm$^{-3}$ giving a similar value of $A \approx 12$~ps$\cdot$K. In order to compare the value $\tau_p = 55$~fs with the previously obtained $\tau_3 = 40$~fs we calculated the ratio of $\tau_p/\tau_3$ by angular integration of the cross-section of partial scattering at the screened Coulomb potential of charged impurities (see Supplemental material \cite{Suppl}). For the parameters of our sample the ratio $\tau_p/\tau_3 = 1.7$ and does not reach the asymptotic value of $6$, obtained for an extremely small scattering angle \cite{Dyakonov1972}. Thus, the time $\tau_3$ obtained by considering classical Dyakonov-Perel relaxation is in good agreement with the time $\tau_p$ derived from the weak localization anomaly.

We have also studied the magnetic field dependencies of $T_1$ and $\rho$ for a sample with higher electron concentration $n_\text{e} = 7.1\times10^{16}$~cm$^{-3}$. The corresponding results are presented in the insets in Figs.~\ref{fig:T1}(a) and \ref{fig:rho}(a) and more in detail in the Supplemental material \cite{Suppl}. One can see that the effect of weak localization is reduced by a factor of about two for $n_\text{e} = 7.1\times10^{16}$~cm$^{-3}$ compared to the sample with $n_\text{e} = 3.7\times10^{16}$~cm$^{-3}$ as expected from Eq.~\eqref{tau:s:b} which contains $n_\text{e}$ in the denominator. The times $\tau_3$, $\tau_p$ and $\tau_\phi$ are similar for both samples.

\emph{In conclusion}, we have demonstrated that the weak localization of electrons has pronounced impact on their spin dynamics. The longitudinal spin relaxation time $T_1$ in $n$-doped GaAs, being in the metallic phase, demonstrates an anomalous decrease with increasing magnetic field at low temperatures. This decrease is due to the field-induced destruction of phase coherence for electrons resulting in the suppression of the weak localization. This shows that physics studied in transport experiments capturing the entirety of physical phenomena between the electrical contacts may be studied locally using focused optical probes of the spin dynamics. The potential of this approach will be very prominent also in two-dimensional systems where one can expect visualization of the weak
localization induced non-exponential tails in spin polarization.

\section{Acknowlegments}
We are grateful to S.~A.~Crooker for providing the samples, and to E.~Evers and A.~Greilich for valuable advice and useful discussions. We acknowledge the financial support of the Deutsche Forschungsgemeinschaft in the frame of the ICRC TRR 160 (project A1). MMG was partially supported by the RFBR Project No. 15-52-12012, Russian Federation President grant MD-1555.2017.2, and the program of RAS ``Nanostructures''. KVK and MMG acknowledge support from Saint-Petersburg State University via a research grant 11.34.2.2012. Magnetotransport measurements were performed using research equipment of
the LPI Shared Facility Center and supported by Ministry of Education and Science of the Russian Federation (Grant No. RFMEFI61717X0001). A.~Yu.~K. was supported by Basic research program of HSE.


\begin{thebibliography}{44}%
\makeatletter
\providecommand \@ifxundefined [1]{%
 \@ifx{#1\undefined}
}%
\providecommand \@ifnum [1]{%
 \ifnum #1\expandafter \@firstoftwo
 \else \expandafter \@secondoftwo
 \fi
}%
\providecommand \@ifx [1]{%
 \ifx #1\expandafter \@firstoftwo
 \else \expandafter \@secondoftwo
 \fi
}%
\providecommand \natexlab [1]{#1}%
\providecommand \enquote  [1]{``#1''}%
\providecommand \bibnamefont  [1]{#1}%
\providecommand \bibfnamefont [1]{#1}%
\providecommand \citenamefont [1]{#1}%
\providecommand \href@noop [0]{\@secondoftwo}%
\providecommand \href [0]{\begingroup \@sanitize@url \@href}%
\providecommand \@href[1]{\@@startlink{#1}\@@href}%
\providecommand \@@href[1]{\endgroup#1\@@endlink}%
\providecommand \@sanitize@url [0]{\catcode `\\12\catcode `\$12\catcode
  `\&12\catcode `\#12\catcode `\^12\catcode `\_12\catcode `\%12\relax}%
\providecommand \@@startlink[1]{}%
\providecommand \@@endlink[0]{}%
\providecommand \url  [0]{\begingroup\@sanitize@url \@url }%
\providecommand \@url [1]{\endgroup\@href {#1}{\urlprefix }}%
\providecommand \urlprefix  [0]{URL }%
\providecommand \Eprint [0]{\href }%
\providecommand \doibase [0]{http://dx.doi.org/}%
\providecommand \selectlanguage [0]{\@gobble}%
\providecommand \bibinfo  [0]{\@secondoftwo}%
\providecommand \bibfield  [0]{\@secondoftwo}%
\providecommand \translation [1]{[#1]}%
\providecommand \BibitemOpen [0]{}%
\providecommand \bibitemStop [0]{}%
\providecommand \bibitemNoStop [0]{.\EOS\space}%
\providecommand \EOS [0]{\spacefactor3000\relax}%
\providecommand \BibitemShut  [1]{\csname bibitem#1\endcsname}%
\let\auto@bib@innerbib\@empty
\bibitem [{\citenamefont {Dzhioev}\ \emph {et~al.}(2002)\citenamefont
  {Dzhioev}, \citenamefont {Kavokin}, \citenamefont {Korenev}, \citenamefont
  {Lazarev}, \citenamefont {Meltser}, \citenamefont {Stepanova}, \citenamefont
  {Zakharchenya}, \citenamefont {Gammon},\ and\ \citenamefont
  {Katzer}}]{Dzhioev2002}%
  \BibitemOpen
  \bibfield  {author} {\bibinfo {author} {\bibfnamefont {R.~I.}\ \bibnamefont
  {Dzhioev}}, \bibinfo {author} {\bibfnamefont {K.~V.}\ \bibnamefont
  {Kavokin}}, \bibinfo {author} {\bibfnamefont {V.~L.}\ \bibnamefont
  {Korenev}}, \bibinfo {author} {\bibfnamefont {M.~V.}\ \bibnamefont
  {Lazarev}}, \bibinfo {author} {\bibfnamefont {B.~Ya.}\ \bibnamefont
  {Meltser}}, \bibinfo {author} {\bibfnamefont {M.~N.}\ \bibnamefont
  {Stepanova}}, \bibinfo {author} {\bibfnamefont {B.~P.}\ \bibnamefont
  {Zakharchenya}}, \bibinfo {author} {\bibfnamefont {D.}~\bibnamefont
  {Gammon}}, \ and\ \bibinfo {author} {\bibfnamefont {D.~S.}\ \bibnamefont
  {Katzer}},\ }\bibfield  {title} {\enquote {\bibinfo {title} {{Low-temperature
  spin relaxation in n-type GaAs}},}\ }\href {\doibase
  10.1103/PhysRevB.66.245204} {\bibfield  {journal} {\bibinfo  {journal} {Phys.
  Rev. B}\ }\textbf {\bibinfo {volume} {66}},\ \bibinfo {pages} {245204}
  (\bibinfo {year} {2002})}\BibitemShut {NoStop}%
\bibitem [{\citenamefont {Belykh}\ \emph {et~al.}(2017)\citenamefont {Belykh},
  \citenamefont {Kavokin}, \citenamefont {Yakovlev},\ and\ \citenamefont
  {Bayer}}]{Belykh2017}%
  \BibitemOpen
  \bibfield  {author} {\bibinfo {author} {\bibfnamefont {V.~V.}\ \bibnamefont
  {Belykh}}, \bibinfo {author} {\bibfnamefont {K.~V.}\ \bibnamefont {Kavokin}},
  \bibinfo {author} {\bibfnamefont {D.~R.}\ \bibnamefont {Yakovlev}}, \ and\
  \bibinfo {author} {\bibfnamefont {M.}~\bibnamefont {Bayer}},\ }\bibfield
  {title} {\enquote {\bibinfo {title} {{Electron charge and spin delocalization
  revealed in the optically probed longitudinal and transverse spin dynamics in
  n-GaAs}},}\ }\href {\doibase 10.1103/PhysRevB.96.241201} {\bibfield
  {journal} {\bibinfo  {journal} {Phys. Rev. B}\ }\textbf {\bibinfo {volume}
  {96}},\ \bibinfo {pages} {241201} (\bibinfo {year} {2017})}\BibitemShut
  {NoStop}%
\bibitem [{\citenamefont {Lonnemann}\ \emph {et~al.}(2017)\citenamefont
  {Lonnemann}, \citenamefont {Rugeramigabo}, \citenamefont {Oestreich},\ and\
  \citenamefont {H\"{u}bner}}]{Lonnemann2017}%
  \BibitemOpen
  \bibfield  {author} {\bibinfo {author} {\bibfnamefont {J.~G.}\ \bibnamefont
  {Lonnemann}}, \bibinfo {author} {\bibfnamefont {E.~P.}\ \bibnamefont
  {Rugeramigabo}}, \bibinfo {author} {\bibfnamefont {M.}~\bibnamefont
  {Oestreich}}, \ and\ \bibinfo {author} {\bibfnamefont {J.}~\bibnamefont
  {H\"{u}bner}},\ }\bibfield  {title} {\enquote {\bibinfo {title} {{Closing the
  gap between spatial and spin dynamics of electrons at the metal-to-insulator
  transition}},}\ }\href {\doibase 10.1103/PhysRevB.96.045201} {\bibfield
  {journal} {\bibinfo  {journal} {Phys. Rev. B}\ }\textbf {\bibinfo {volume}
  {96}},\ \bibinfo {pages} {045201} (\bibinfo {year} {2017})}\BibitemShut
  {NoStop}%
\bibitem [{\citenamefont {D’yakonov}\ and\ \citenamefont
  {Perel’}(1972)}]{Dyakonov1972}%
  \BibitemOpen
  \bibfield  {author} {\bibinfo {author} {\bibfnamefont {M.~I.}\ \bibnamefont
  {D’yakonov}}\ and\ \bibinfo {author} {\bibfnamefont {V.~I.}\ \bibnamefont
  {Perel’}},\ }\bibfield  {title} {\enquote {\bibinfo {title} {{Spin relaxation of conduction electrons in
  noncentrosymetric semiconductors}},}\ }\href@noop {} {\bibfield  {journal}
  {\bibinfo  {journal} {Fiz. Tverd. Tela}\ }\textbf {\bibinfo {volume} {13}},\
  \bibinfo {pages} {3581} (\bibinfo {year} {1972}) [Sov. Phys.
  Solid State 13, 3023 (1972)]}\BibitemShut {NoStop}%
\bibitem [{\citenamefont {Kavokin}(2008)}]{Kavokin08}%
  \BibitemOpen
  \bibfield  {author} {\bibinfo {author} {\bibfnamefont {K.~V.}\ \bibnamefont
  {Kavokin}},\ }\bibfield  {title} {\enquote {\bibinfo {title} {{Spin
  relaxation of localized electrons in n-type semiconductors}},}\ }\href
  {\doibase 10.1088/0268-1242/23/11/114009} {\bibfield  {journal} {\bibinfo
  {journal} {Semicond. Sci. Technol.}\ }\textbf {\bibinfo {volume} {23}},\
  \bibinfo {pages} {114009} (\bibinfo {year} {2008})}\BibitemShut {NoStop}%
\bibitem [{\citenamefont {Dyakonov}\ and\ \citenamefont
  {Kachorovskii}(1986)}]{Dyakonov86}%
  \BibitemOpen
  \bibfield  {author} {\bibinfo {author} {\bibfnamefont {M.}~\bibnamefont
  {Dyakonov}}\ and\ \bibinfo {author} {\bibfnamefont {V.}~\bibnamefont
  {Kachorovskii}},\ }\bibfield  {title} {\enquote {\bibinfo {title} {{Spin
  relaxation of two-dimensional electrons in noncentrosymmetric
  semiconductors}},}\ }\href@noop {} {\bibfield  {journal} {\bibinfo  {journal}
  {Sov. Phys. Semicond.}\ }\textbf {\bibinfo {volume} {20}},\ \bibinfo {pages}
  {110} (\bibinfo {year} {1986})}\BibitemShut {NoStop}%
\bibitem [{\citenamefont {Dyakonov}(2017)}]{Dyakonov2017}%
  \BibitemOpen
  \bibinfo {editor} {\bibfnamefont {M.~I.}\ \bibnamefont {Dyakonov}},\
  ed.,\ \href {\doibase 10.1007/978-3-319-65436-2} {\emph {\bibinfo {title}
  {{Spin Physics in Semiconductors}}}},\ \bibinfo {series} {Springer Series in
  Solid-State Sciences}, Vol.\ \bibinfo {volume} {157}\ (\bibinfo  {publisher}
  {Springer International Publishing},\ \bibinfo {address} {Cham},\ \bibinfo
  {year} {2017})\BibitemShut {NoStop}%
\bibitem [{\citenamefont {Sih}\ \emph {et~al.}(2004)\citenamefont {Sih},
  \citenamefont {Lau}, \citenamefont {Myers}, \citenamefont {Gossard},
  \citenamefont {Flatt\'{e}},\ and\ \citenamefont {Awschalom}}]{Sih2004}%
  \BibitemOpen
  \bibfield  {author} {\bibinfo {author} {\bibfnamefont {V.}~\bibnamefont
  {Sih}}, \bibinfo {author} {\bibfnamefont {W.~H.}\ \bibnamefont {Lau}},
  \bibinfo {author} {\bibfnamefont {R.~C.}\ \bibnamefont {Myers}}, \bibinfo
  {author} {\bibfnamefont {A.~C.}\ \bibnamefont {Gossard}}, \bibinfo {author}
  {\bibfnamefont {M.~E.}\ \bibnamefont {Flatt\'{e}}}, \ and\ \bibinfo {author}
  {\bibfnamefont {D.~D.}\ \bibnamefont {Awschalom}},\ }\bibfield  {title}
  {\enquote {\bibinfo {title} {{Control of electron-spin coherence using Landau
  level quantization in a two-dimensional electron gas}},}\ }\href {\doibase
  10.1103/PhysRevB.70.161313} {\bibfield  {journal} {\bibinfo  {journal} {Phys.
  Rev. B}\ }\textbf {\bibinfo {volume} {70}},\ \bibinfo {pages} {161313}
  (\bibinfo {year} {2004})}\BibitemShut {NoStop}%
\bibitem [{\citenamefont {Fukuoka}\ \emph {et~al.}(2008)\citenamefont
  {Fukuoka}, \citenamefont {Yamazaki}, \citenamefont {Tanaka}, \citenamefont
  {Oto}, \citenamefont {Muro}, \citenamefont {Hirayama}, \citenamefont
  {Kumada},\ and\ \citenamefont {Yamaguchi}}]{Fukuoka2008}%
  \BibitemOpen
  \bibfield  {author} {\bibinfo {author} {\bibfnamefont {D.}~\bibnamefont
  {Fukuoka}}, \bibinfo {author} {\bibfnamefont {T.}~\bibnamefont {Yamazaki}},
  \bibinfo {author} {\bibfnamefont {N.}~\bibnamefont {Tanaka}}, \bibinfo
  {author} {\bibfnamefont {K.}~\bibnamefont {Oto}}, \bibinfo {author}
  {\bibfnamefont {K.}~\bibnamefont {Muro}}, \bibinfo {author} {\bibfnamefont
  {Y.}~\bibnamefont {Hirayama}}, \bibinfo {author} {\bibfnamefont
  {N.}~\bibnamefont {Kumada}}, \ and\ \bibinfo {author} {\bibfnamefont
  {H.}~\bibnamefont {Yamaguchi}},\ }\bibfield  {title} {\enquote {\bibinfo
  {title} {{Spin dynamics of two-dimensional electrons in a quantum Hall system
  probed by time-resolved Kerr rotation spectroscopy}},}\ }\href {\doibase
  10.1103/PhysRevB.78.041304} {\bibfield  {journal} {\bibinfo  {journal} {Phys.
  Rev. B}\ }\textbf {\bibinfo {volume} {78}},\ \bibinfo {pages} {041304}
  (\bibinfo {year} {2008})}\BibitemShut {NoStop}%
\bibitem [{\citenamefont {Larionov}\ \emph {et~al.}(2015)\citenamefont
  {Larionov}, \citenamefont {Kulik}, \citenamefont {Dickmann},\ and\
  \citenamefont {Kukushkin}}]{Larionov2015}%
  \BibitemOpen
  \bibfield  {author} {\bibinfo {author} {\bibfnamefont {A.~V.}\ \bibnamefont
  {Larionov}}, \bibinfo {author} {\bibfnamefont {L.~V.}\ \bibnamefont {Kulik}},
  \bibinfo {author} {\bibfnamefont {S.}~\bibnamefont {Dickmann}}, \ and\
  \bibinfo {author} {\bibfnamefont {I.~V.}\ \bibnamefont {Kukushkin}},\
  }\bibfield  {title} {\enquote {\bibinfo {title} {{Goldstone mode
  stochastization in a quantum Hall ferromagnet}},}\ }\href {\doibase
  10.1103/PhysRevB.92.165417} {\bibfield  {journal} {\bibinfo  {journal} {Phys.
  Rev. B}\ }\textbf {\bibinfo {volume} {92}},\ \bibinfo {pages} {165417}
  (\bibinfo {year} {2015})}\BibitemShut {NoStop}%
\bibitem [{\citenamefont {Larionov}\ \emph {et~al.}(2017)\citenamefont
  {Larionov}, \citenamefont {Stepanets-Khussein},\ and\ \citenamefont
  {Kulik}}]{Larionov2017}%
  \BibitemOpen
  \bibfield  {author} {\bibinfo {author} {\bibfnamefont {A.~V.}\ \bibnamefont
  {Larionov}}, \bibinfo {author} {\bibfnamefont {E.}~\bibnamefont
  {Stepanets-Khussein}}, \ and\ \bibinfo {author} {\bibfnamefont {L.~V.}\
  \bibnamefont {Kulik}},\ }\bibfield  {title} {\enquote {\bibinfo {title}
  {{Spin dephasing of a two-dimensional electron gas in a GaAs quantum well
  near odd filling factors}},}\ }\href {\doibase 10.1134/S0021364017040129}
  {\bibfield  {journal} {\bibinfo  {journal} {JETP Lett.}\ }\textbf {\bibinfo
  {volume} {105}},\ \bibinfo {pages} {238} (\bibinfo {year}
  {2017})}\BibitemShut {NoStop}%
\bibitem [{\citenamefont {Shklovskii}(2006)}]{Shklovskii06}%
  \BibitemOpen
  \bibfield  {author} {\bibinfo {author} {\bibfnamefont {B.~I.}\ \bibnamefont
  {Shklovskii}},\ }\bibfield  {title} {\enquote {\bibinfo {title}
  {{Dyakonov-Perel spin relaxation near the metal-insulator transition and in
  hopping transport}},}\ }\href {\doibase 10.1103/PhysRevB.73.193201}
  {\bibfield  {journal} {\bibinfo  {journal} {Phys. Rev. B}\ }\textbf {\bibinfo
  {volume} {73}},\ \bibinfo {pages} {193201} (\bibinfo {year}
  {2006})}\BibitemShut {NoStop}%
\bibitem [{\citenamefont {Lyubinskiy}\ and\ \citenamefont
  {Kachorovskii}(2004)}]{Lyubinskiy2004}%
  \BibitemOpen
  \bibfield  {author} {\bibinfo {author} {\bibfnamefont {I.~S.}\ \bibnamefont
  {Lyubinskiy}}\ and\ \bibinfo {author} {\bibfnamefont {V.~Yu.}\ \bibnamefont
  {Kachorovskii}},\ }\bibfield  {title} {\enquote {\bibinfo {title} {{Slowing
  down of spin relaxation in two-dimensional systems by quantum interference
  effects}},}\ }\href {\doibase 10.1103/PhysRevB.70.205335} {\bibfield
  {journal} {\bibinfo  {journal} {Phys. Rev. B}\ }\textbf {\bibinfo {volume}
  {70}},\ \bibinfo {pages} {205335} (\bibinfo {year} {2004})}\BibitemShut
  {NoStop}%
\bibitem [{\citenamefont {Pikus}\ and\ \citenamefont
  {Titkov}(1984)}]{Pikus1984}%
  \BibitemOpen
  \bibfield  {author} {\bibinfo {author} {\bibfnamefont {G.~E.}\ \bibnamefont
  {Pikus}}\ and\ \bibinfo {author} {\bibfnamefont {A.~N.}\ \bibnamefont
  {Titkov}},\ }\bibfield  {title} {\enquote {\bibinfo {title} {{Spin relaxation
  under optical orientation in semiconductors}},}\ }in\ \href@noop {} {\emph
  {\bibinfo {booktitle} {Optical Orientation}}},\ \bibinfo {editor} {edited by\
  \bibinfo {editor} {\bibfnamefont {F.}~\bibnamefont {Meier}}\ and\ \bibinfo
  {editor} {\bibfnamefont {B.~P.}\ \bibnamefont {Zakharchenya}}}\ (\bibinfo
  {publisher} {North-Holland},\ \bibinfo {address} {Amsterdam},\ \bibinfo
  {year} {1984})\ \BibitemShut {NoStop}%
\bibitem [{\citenamefont {Baumberg}\ \emph {et~al.}(1994)\citenamefont
  {Baumberg}, \citenamefont {Awschalom}, \citenamefont {Samarth}, \citenamefont
  {Luo},\ and\ \citenamefont {Furdyna}}]{Baumberg1994}%
  \BibitemOpen
  \bibfield  {author} {\bibinfo {author} {\bibfnamefont {J.~J.}\ \bibnamefont
  {Baumberg}}, \bibinfo {author} {\bibfnamefont {D.~D.}\ \bibnamefont
  {Awschalom}}, \bibinfo {author} {\bibfnamefont {N.}~\bibnamefont {Samarth}},
  \bibinfo {author} {\bibfnamefont {H.}~\bibnamefont {Luo}}, \ and\ \bibinfo
  {author} {\bibfnamefont {J.~K.}\ \bibnamefont {Furdyna}},\ }\bibfield
  {title} {\enquote {\bibinfo {title} {{Spin beats and dynamical magnetization
  in quantum structures}},}\ }\href {\doibase 10.1103/PhysRevLett.72.717}
  {\bibfield  {journal} {\bibinfo  {journal} {Phys. Rev. Lett.}\ }\textbf
  {\bibinfo {volume} {72}},\ \bibinfo {pages} {717} (\bibinfo {year}
  {1994})}\BibitemShut {NoStop}%
\bibitem [{\citenamefont {Zheludev}\ \emph {et~al.}(1994)\citenamefont
  {Zheludev}, \citenamefont {Brummell}, \citenamefont {Harley}, \citenamefont
  {Malinowski}, \citenamefont {Popov}, \citenamefont {Ashenford},\ and\
  \citenamefont {Lunn}}]{Zheludev1994}%
  \BibitemOpen
  \bibfield  {author} {\bibinfo {author} {\bibfnamefont {N.I.}\ \bibnamefont
  {Zheludev}}, \bibinfo {author} {\bibfnamefont {M.A.}\ \bibnamefont
  {Brummell}}, \bibinfo {author} {\bibfnamefont {R.T.}\ \bibnamefont {Harley}},
  \bibinfo {author} {\bibfnamefont {A.}~\bibnamefont {Malinowski}}, \bibinfo
  {author} {\bibfnamefont {S.V.}\ \bibnamefont {Popov}}, \bibinfo {author}
  {\bibfnamefont {D.E.}\ \bibnamefont {Ashenford}}, \ and\ \bibinfo {author}
  {\bibfnamefont {B.}~\bibnamefont {Lunn}},\ }\bibfield  {title} {\enquote
  {\bibinfo {title} {{Giant specular inverse Faraday effect in
  Cd0.6Mn0.4Te}},}\ }\href {\doibase 10.1016/0038-1098(94)90064-7} {\bibfield
  {journal} {\bibinfo  {journal} {Solid State Commun.}\ }\textbf {\bibinfo
  {volume} {89}},\ \bibinfo {pages} {823} (\bibinfo {year}
  {1994})}\BibitemShut {NoStop}%
\bibitem [{\citenamefont {Colton}\ \emph {et~al.}(2004)\citenamefont {Colton},
  \citenamefont {Kennedy}, \citenamefont {Bracker},\ and\ \citenamefont
  {Gammon}}]{Colton2004}%
  \BibitemOpen
  \bibfield  {author} {\bibinfo {author} {\bibfnamefont {J.~S.}\ \bibnamefont
  {Colton}}, \bibinfo {author} {\bibfnamefont {T.~A.}\ \bibnamefont {Kennedy}},
  \bibinfo {author} {\bibfnamefont {A.~S.}\ \bibnamefont {Bracker}}, \ and\
  \bibinfo {author} {\bibfnamefont {D.}~\bibnamefont {Gammon}},\ }\bibfield
  {title} {\enquote {\bibinfo {title} {{Microsecond spin-flip times in n-GaAs
  measured by time-resolved polarization of photoluminescence}},}\ }\href
  {\doibase 10.1103/PhysRevB.69.121307} {\bibfield  {journal} {\bibinfo
  {journal} {Phys. Rev. B}\ }\textbf {\bibinfo {volume} {69}},\ \bibinfo
  {pages} {121307} (\bibinfo {year} {2004})}\BibitemShut {NoStop}%
\bibitem [{\citenamefont {Colton}\ \emph {et~al.}(2007)\citenamefont {Colton},
  \citenamefont {Heeb}, \citenamefont {Schroeder}, \citenamefont {Stokes},
  \citenamefont {Wienkes},\ and\ \citenamefont {Bracker}}]{Colton2007}%
  \BibitemOpen
  \bibfield  {author} {\bibinfo {author} {\bibfnamefont {J.~S.}\ \bibnamefont
  {Colton}}, \bibinfo {author} {\bibfnamefont {M.~E.}\ \bibnamefont {Heeb}},
  \bibinfo {author} {\bibfnamefont {P.}~\bibnamefont {Schroeder}}, \bibinfo
  {author} {\bibfnamefont {A.}~\bibnamefont {Stokes}}, \bibinfo {author}
  {\bibfnamefont {L.~R.}\ \bibnamefont {Wienkes}}, \ and\ \bibinfo {author}
  {\bibfnamefont {A.~S.}\ \bibnamefont {Bracker}},\ }\bibfield  {title}
  {\enquote {\bibinfo {title} {{Anomalous magnetic field dependence of the T1
  spin lifetime in a lightly doped GaAs sample}},}\ }\href {\doibase
  10.1103/PhysRevB.75.205201} {\bibfield  {journal} {\bibinfo  {journal} {Phys.
  Rev. B}\ }\textbf {\bibinfo {volume} {75}},\ \bibinfo {pages} {205201}
  (\bibinfo {year} {2007})}\BibitemShut {NoStop}%
\bibitem [{\citenamefont {Fu}\ \emph {et~al.}(2006)\citenamefont {Fu},
  \citenamefont {Yeo}, \citenamefont {Clark}, \citenamefont {Santori},
  \citenamefont {Stanley}, \citenamefont {Holland},\ and\ \citenamefont
  {Yamamoto}}]{Fu2006}%
  \BibitemOpen
  \bibfield  {author} {\bibinfo {author} {\bibfnamefont {Kai-Mei~C.}\
  \bibnamefont {Fu}}, \bibinfo {author} {\bibfnamefont {Wenzheng}\ \bibnamefont
  {Yeo}}, \bibinfo {author} {\bibfnamefont {Susan}\ \bibnamefont {Clark}},
  \bibinfo {author} {\bibfnamefont {Charles}\ \bibnamefont {Santori}}, \bibinfo
  {author} {\bibfnamefont {Colin}\ \bibnamefont {Stanley}}, \bibinfo {author}
  {\bibfnamefont {M.~C.}\ \bibnamefont {Holland}}, \ and\ \bibinfo {author}
  {\bibfnamefont {Yoshihisa}\ \bibnamefont {Yamamoto}},\ }\bibfield  {title}
  {\enquote {\bibinfo {title} {{Millisecond spin-flip times of donor-bound
  electrons in GaAs}},}\ }\href {\doibase 10.1103/PhysRevB.74.121304}
  {\bibfield  {journal} {\bibinfo  {journal} {Phys. Rev. B}\ }\textbf {\bibinfo
  {volume} {74}},\ \bibinfo {pages} {121304} (\bibinfo {year}
  {2006})}\BibitemShut {NoStop}%
\bibitem [{\citenamefont {Linpeng}\ \emph {et~al.}(2016)\citenamefont
  {Linpeng}, \citenamefont {Karin}, \citenamefont {Durnev}, \citenamefont
  {Barbour}, \citenamefont {Glazov}, \citenamefont {Sherman}, \citenamefont
  {Watkins}, \citenamefont {Seto},\ and\ \citenamefont {Fu}}]{Linpeng2016}%
  \BibitemOpen
  \bibfield  {author} {\bibinfo {author} {\bibfnamefont {X.}\ \bibnamefont
  {Linpeng}}, \bibinfo {author} {\bibfnamefont {T.}\ \bibnamefont {Karin}},
  \bibinfo {author} {\bibfnamefont {M.~V.}\ \bibnamefont {Durnev}}, \bibinfo
  {author} {\bibfnamefont {R.}\ \bibnamefont {Barbour}}, \bibinfo {author}
  {\bibfnamefont {M.~M.}\ \bibnamefont {Glazov}}, \bibinfo {author}
  {\bibfnamefont {E.~Ya.}\ \bibnamefont {Sherman}}, \bibinfo {author}
  {\bibfnamefont {S.~P.}\ \bibnamefont {Watkins}}, \bibinfo {author}
  {\bibfnamefont {S.}\ \bibnamefont {Seto}}, \ and\ \bibinfo {author}
  {\bibfnamefont {Kai-Mei~C.}\ \bibnamefont {Fu}},\ }\bibfield  {title}
  {\enquote {\bibinfo {title} {{Longitudinal spin relaxation of donor-bound
  electrons in direct band-gap semiconductors}},}\ }\href {\doibase
  10.1103/PhysRevB.94.125401} {\bibfield  {journal} {\bibinfo  {journal} {Phys.
  Rev. B}\ }\textbf {\bibinfo {volume} {94}},\ \bibinfo {pages} {125401}
  (\bibinfo {year} {2016})}\BibitemShut {NoStop}%
\bibitem [{\citenamefont {Kikkawa}\ and\ \citenamefont
  {Awschalom}(1998)}]{Kikkawa1998}%
  \BibitemOpen
  \bibfield  {author} {\bibinfo {author} {\bibfnamefont {J.~M.}\ \bibnamefont
  {Kikkawa}}\ and\ \bibinfo {author} {\bibfnamefont {D.~D.}\ \bibnamefont
  {Awschalom}},\ }\bibfield  {title} {\enquote {\bibinfo {title} {{Resonant
  Spin Amplification in n-Type GaAs}},}\ }\href {\doibase
  10.1103/PhysRevLett.80.4313} {\bibfield  {journal} {\bibinfo  {journal}
  {Phys. Rev. Lett.}\ }\textbf {\bibinfo {volume} {80}},\ \bibinfo {pages}
  {4313} (\bibinfo {year} {1998})}\BibitemShut {NoStop}%
\bibitem [{\citenamefont {Yugova}\ \emph {et~al.}(2012)\citenamefont {Yugova},
  \citenamefont {Glazov}, \citenamefont {Yakovlev}, \citenamefont {Sokolova},\
  and\ \citenamefont {Bayer}}]{Yugova2012}%
  \BibitemOpen
  \bibfield  {author} {\bibinfo {author} {\bibfnamefont {I.~A.}\ \bibnamefont
  {Yugova}}, \bibinfo {author} {\bibfnamefont {M.~M.}\ \bibnamefont {Glazov}},
  \bibinfo {author} {\bibfnamefont {D.~R.}\ \bibnamefont {Yakovlev}}, \bibinfo
  {author} {\bibfnamefont {A.~A.}\ \bibnamefont {Sokolova}}, \ and\ \bibinfo
  {author} {\bibfnamefont {M.}~\bibnamefont {Bayer}},\ }\bibfield  {title}
  {\enquote {\bibinfo {title} {{Coherent spin dynamics of electrons and holes
  in semiconductor quantum wells and quantum dots under periodical optical
  excitation: Resonant spin amplification versus spin mode locking}},}\ }\href
  {\doibase 10.1103/PhysRevB.85.125304} {\bibfield  {journal} {\bibinfo
  {journal} {Phys. Rev. B}\ }\textbf {\bibinfo {volume} {85}},\ \bibinfo
  {pages} {125304} (\bibinfo {year} {2012})}\BibitemShut {NoStop}%
\bibitem [{\citenamefont {Oestreich}\ \emph {et~al.}(2005)\citenamefont
  {Oestreich}, \citenamefont {R\"{o}mer}, \citenamefont {Haug},\ and\
  \citenamefont {H\"{a}gele}}]{Oestreich2005}%
  \BibitemOpen
  \bibfield  {author} {\bibinfo {author} {\bibfnamefont {M.}~\bibnamefont
  {Oestreich}}, \bibinfo {author} {\bibfnamefont {M.}~\bibnamefont
  {R\"{o}mer}}, \bibinfo {author} {\bibfnamefont {R.~J.}\ \bibnamefont {Haug}},
  \ and\ \bibinfo {author} {\bibfnamefont {D.}~\bibnamefont {H\"{a}gele}},\
  }\bibfield  {title} {\enquote {\bibinfo {title} {{Spin Noise Spectroscopy in
  GaAs}},}\ }\href {\doibase 10.1103/PhysRevLett.95.216603} {\bibfield
  {journal} {\bibinfo  {journal} {Phys. Rev. Lett.}\ }\textbf {\bibinfo
  {volume} {95}},\ \bibinfo {pages} {216603} (\bibinfo {year}
  {2005})}\BibitemShut {NoStop}%
\bibitem [{\citenamefont {Crooker}\ \emph {et~al.}(2009)\citenamefont
  {Crooker}, \citenamefont {Cheng},\ and\ \citenamefont {Smith}}]{Crooker2009}%
  \BibitemOpen
  \bibfield  {author} {\bibinfo {author} {\bibfnamefont {S.~A.}\
  \bibnamefont {Crooker}}, \bibinfo {author} {\bibfnamefont {L.}\
  \bibnamefont {Cheng}}, \ and\ \bibinfo {author} {\bibfnamefont {D.~L.}\
  \bibnamefont {Smith}},\ }\bibfield  {title} {\enquote {\bibinfo {title}
  {{Spin noise of conduction electrons in n-type bulk GaAs}},}\ }\href
  {\doibase 10.1103/PhysRevB.79.035208} {\bibfield  {journal} {\bibinfo
  {journal} {Phys. Rev. B}\ }\textbf {\bibinfo {volume} {79}},\ \bibinfo
  {pages} {035208} (\bibinfo {year} {2009})}\BibitemShut {NoStop}%
\bibitem [{\citenamefont {R\"{o}mer}\ \emph {et~al.}(2010)\citenamefont
  {R\"{o}mer}, \citenamefont {Bernien}, \citenamefont {M\"{u}ller},
  \citenamefont {Schuh}, \citenamefont {H\"{u}bner},\ and\ \citenamefont
  {Oestreich}}]{Romer2010}%
  \BibitemOpen
  \bibfield  {author} {\bibinfo {author} {\bibfnamefont {M.}~\bibnamefont
  {R\"{o}mer}}, \bibinfo {author} {\bibfnamefont {H.}~\bibnamefont {Bernien}},
  \bibinfo {author} {\bibfnamefont {G.}~\bibnamefont {M\"{u}ller}}, \bibinfo
  {author} {\bibfnamefont {D.}~\bibnamefont {Schuh}}, \bibinfo {author}
  {\bibfnamefont {J.}~\bibnamefont {H\"{u}bner}}, \ and\ \bibinfo {author}
  {\bibfnamefont {M.}~\bibnamefont {Oestreich}},\ }\bibfield  {title} {\enquote
  {\bibinfo {title} {{Electron-spin relaxation in bulk GaAs for doping
  densities close to the metal-to-insulator transition}},}\ }\href {\doibase
  10.1103/PhysRevB.81.075216} {\bibfield  {journal} {\bibinfo  {journal} {Phys.
  Rev. B}\ }\textbf {\bibinfo {volume} {81}},\ \bibinfo {pages} {075216}
  (\bibinfo {year} {2010})}\BibitemShut {NoStop}%
\bibitem [{\citenamefont {Heisterkamp}\ \emph {et~al.}(2015)\citenamefont
  {Heisterkamp}, \citenamefont {Zhukov}, \citenamefont {Greilich},
  \citenamefont {Yakovlev}, \citenamefont {Korenev}, \citenamefont {Pawlis},\
  and\ \citenamefont {Bayer}}]{Heisterkamp2015}%
  \BibitemOpen
  \bibfield  {author} {\bibinfo {author} {\bibfnamefont {F.}~\bibnamefont
  {Heisterkamp}}, \bibinfo {author} {\bibfnamefont {E.~A.}\ \bibnamefont
  {Zhukov}}, \bibinfo {author} {\bibfnamefont {A.}~\bibnamefont {Greilich}},
  \bibinfo {author} {\bibfnamefont {D.~R.}\ \bibnamefont {Yakovlev}}, \bibinfo
  {author} {\bibfnamefont {V.~L.}\ \bibnamefont {Korenev}}, \bibinfo {author}
  {\bibfnamefont {A.}~\bibnamefont {Pawlis}}, \ and\ \bibinfo {author}
  {\bibfnamefont {M.}~\bibnamefont {Bayer}},\ }\bibfield  {title} {\enquote
  {\bibinfo {title} {{Longitudinal and transverse spin dynamics of donor-bound
  electrons in fluorine-doped ZnSe: Spin inertia versus Hanle effect}},}\
  }\href {\doibase 10.1103/PhysRevB.91.235432} {\bibfield  {journal} {\bibinfo
  {journal} {Phys. Rev. B}\ }\textbf {\bibinfo {volume} {91}},\ \bibinfo
  {pages} {235432} (\bibinfo {year} {2015})}\BibitemShut {NoStop}%
\bibitem [{\citenamefont {Belykh}\ \emph {et~al.}(2016)\citenamefont {Belykh},
  \citenamefont {Evers}, \citenamefont {Yakovlev}, \citenamefont {Fobbe},
  \citenamefont {Greilich},\ and\ \citenamefont {Bayer}}]{Belykh2016}%
  \BibitemOpen
  \bibfield  {author} {\bibinfo {author} {\bibfnamefont {V.~V.}\ \bibnamefont
  {Belykh}}, \bibinfo {author} {\bibfnamefont {E.}~\bibnamefont {Evers}},
  \bibinfo {author} {\bibfnamefont {D.~R.}\ \bibnamefont {Yakovlev}}, \bibinfo
  {author} {\bibfnamefont {F.}~\bibnamefont {Fobbe}}, \bibinfo {author}
  {\bibfnamefont {A.}~\bibnamefont {Greilich}}, \ and\ \bibinfo {author}
  {\bibfnamefont {M.}~\bibnamefont {Bayer}},\ }\bibfield  {title} {\enquote
  {\bibinfo {title} {{Extended pump-probe Faraday rotation spectroscopy of the
  submicrosecond electron spin dynamics in n-type GaAs}},}\ }\href {\doibase
  10.1103/PhysRevB.94.241202} {\bibfield  {journal} {\bibinfo  {journal} {Phys.
  Rev. B}\ }\textbf {\bibinfo {volume} {94}},\ \bibinfo {pages} {241202}
  (\bibinfo {year} {2016})}\BibitemShut {NoStop}%
\bibitem [{\citenamefont {Altshuler}\ and\ \citenamefont
  {Aronov}(1985)}]{Altshuler1985}%
  \BibitemOpen
  \bibfield  {author} {\bibinfo {author} {\bibfnamefont {B.~L.}\ \bibnamefont
  {Altshuler}}\ and\ \bibinfo {author} {\bibfnamefont {A.~G.}\ \bibnamefont
  {Aronov}},\ }\href@noop {} {\emph {\bibinfo {title} {{Electron-Electron
  Interactions in Disordered Systems}}}},\ edited by\ \bibinfo {editor}
  {\bibfnamefont {A.~L.}\ \bibnamefont {Efros}}\ and\ \bibinfo {editor}
  {\bibfnamefont {M.}~\bibnamefont {Pollak}}\ (\bibinfo  {publisher}
  {Elsevier},\ \bibinfo {address} {Amsterdam},\ \bibinfo {year}
  {1985})\BibitemShut {NoStop}%
\bibitem [{\citenamefont {Fritzsche}\ and\ \citenamefont
  {Lark-Horovitz}(1955)}]{Fritzsche1955}%
  \BibitemOpen
  \bibfield  {author} {\bibinfo {author} {\bibfnamefont {H.}~\bibnamefont
  {Fritzsche}}\ and\ \bibinfo {author} {\bibfnamefont {K.}~\bibnamefont
  {Lark-Horovitz}},\ }\bibfield  {title} {\enquote {\bibinfo {title}
  {{Electrical Properties of p-Type Indium Antimonide at Low Temperatures}},}\
  }\href {\doibase 10.1103/PhysRev.99.400} {\bibfield  {journal} {\bibinfo
  {journal} {Phys. Rev.}\ }\textbf {\bibinfo {volume} {99}},\ \bibinfo {pages}
  {400} (\bibinfo {year} {1955})}\BibitemShut {NoStop}%
\bibitem [{\citenamefont {Woods}\ and\ \citenamefont {Chen}(1964)}]{Woods1964}%
  \BibitemOpen
  \bibfield  {author} {\bibinfo {author} {\bibfnamefont {J.~F.}\ \bibnamefont
  {Woods}}\ and\ \bibinfo {author} {\bibfnamefont {C.~Y.}\ \bibnamefont
  {Chen}},\ }\bibfield  {title} {\enquote {\bibinfo {title} {{Negative
  Magnetoresistance in Impurity Conduction}},}\ }\href {\doibase
  10.1103/PhysRev.135.A1462} {\bibfield  {journal} {\bibinfo  {journal} {Phys.
  Rev.}\ }\textbf {\bibinfo {volume} {135}},\ \bibinfo {pages} {A1462}
  (\bibinfo {year} {1964})}\BibitemShut {NoStop}%
\bibitem [{\citenamefont {Halbo}\ and\ \citenamefont
  {Sladek}(1968)}]{Halbo1968}%
  \BibitemOpen
  \bibfield  {author} {\bibinfo {author} {\bibfnamefont {L.}~\bibnamefont
  {Halbo}}\ and\ \bibinfo {author} {\bibfnamefont {R.~J.}\ \bibnamefont
  {Sladek}},\ }\bibfield  {title} {\enquote {\bibinfo {title}
  {{Magnetoresistance of Undoped n-Type Gallium Arsenide at Low
  Temperatures}},}\ }\href {\doibase 10.1103/PhysRev.173.794} {\bibfield
  {journal} {\bibinfo  {journal} {Phys. Rev.}\ }\textbf {\bibinfo {volume}
  {173}},\ \bibinfo {pages} {794} (\bibinfo {year} {1968})}\BibitemShut
  {NoStop}%
\bibitem [{\citenamefont {Benzaquen}\ \emph {et~al.}(1988)\citenamefont
  {Benzaquen}, \citenamefont {Walsh},\ and\ \citenamefont
  {Mazuruk}}]{Benzaquen1988}%
  \BibitemOpen
  \bibfield  {author} {\bibinfo {author} {\bibfnamefont {M.}~\bibnamefont
  {Benzaquen}}, \bibinfo {author} {\bibfnamefont {D.}~\bibnamefont {Walsh}}, \
  and\ \bibinfo {author} {\bibfnamefont {K.}~\bibnamefont {Mazuruk}},\
  }\bibfield  {title} {\enquote {\bibinfo {title} {{Low-field magnetoresistance
  of n-type GaAs in the variable-range hopping regime}},}\ }\href {\doibase
  10.1103/PhysRevB.38.10933} {\bibfield  {journal} {\bibinfo  {journal} {Phys.
  Rev. B}\ }\textbf {\bibinfo {volume} {38}},\ \bibinfo {pages} {10933}
  (\bibinfo {year} {1988})}\BibitemShut {NoStop}%
\bibitem [{\citenamefont {Kawabata}(1980{\natexlab{a}})}]{Kawabata1980}%
  \BibitemOpen
  \bibfield  {author} {\bibinfo {author} {\bibfnamefont {A.}~\bibnamefont
  {Kawabata}},\ }\bibfield  {title} {\enquote {\bibinfo {title} {{Theory of
  negative magnetoresistance in three-dimensional systems}},}\ }\href {\doibase
  10.1016/0038-1098(80)90644-4} {\bibfield  {journal} {\bibinfo  {journal}
  {Solid State Commun.}\ }\textbf {\bibinfo {volume} {34}},\ \bibinfo {pages}
  {431} (\bibinfo {year} {1980}{\natexlab{a}})}\BibitemShut {NoStop}%
\bibitem [{\citenamefont {Kawabata}(1980{\natexlab{b}})}]{Kawabata1980a}%
  \BibitemOpen
  \bibfield  {author} {\bibinfo {author} {\bibfnamefont {A.}\ \bibnamefont
  {Kawabata}},\ }\bibfield  {title} {\enquote {\bibinfo {title} {{Theory of
  Negative Magnetoresistance I. Application to Heavily Doped
  Semiconductors}},}\ }\href {\doibase 10.1143/JPSJ.49.628} {\bibfield
  {journal} {\bibinfo  {journal} {J. Phys. Soc. Japan}\ }\textbf {\bibinfo
  {volume} {49}},\ \bibinfo {pages} {628} (\bibinfo {year}
  {1980}{\natexlab{b}})}\BibitemShut {NoStop}%
\bibitem [{\citenamefont {Capoen}\ \emph {et~al.}(1993)\citenamefont {Capoen},
  \citenamefont {Biskupski},\ and\ \citenamefont {Briggs}}]{Capoen1993}%
  \BibitemOpen
  \bibfield  {author} {\bibinfo {author} {\bibfnamefont {B}~\bibnamefont
  {Capoen}}, \bibinfo {author} {\bibfnamefont {G}~\bibnamefont {Biskupski}}, \
  and\ \bibinfo {author} {\bibfnamefont {A}~\bibnamefont {Briggs}},\ }\bibfield
   {title} {\enquote {\bibinfo {title} {{Low-temperature conductivity and
  weak-localization effect in barely metallic GaAs}},}\ }\href {\doibase
  10.1088/0953-8984/5/16/012} {\bibfield  {journal} {\bibinfo  {journal} {J.
  Phys. Condens. Matter}\ }\textbf {\bibinfo {volume} {5}},\ \bibinfo {pages}
  {2545} (\bibinfo {year} {1993})}\BibitemShut {NoStop}%
\bibitem [{\citenamefont {Lyubinskiy}\ and\ \citenamefont
  {Kachorovskii}(2005)}]{Lyubinskiy2005}%
  \BibitemOpen
  \bibfield  {author} {\bibinfo {author} {\bibfnamefont {I.~S.}\ \bibnamefont
  {Lyubinskiy}}\ and\ \bibinfo {author} {\bibfnamefont {V.~Yu.}\ \bibnamefont
  {Kachorovskii}},\ }\bibfield  {title} {\enquote {\bibinfo {title} {{Hanle
  Effect Driven by Weak Localization}},}\ }\href {\doibase
  10.1103/PhysRevLett.94.076406} {\bibfield  {journal} {\bibinfo  {journal}
  {Phys. Rev. Lett.}\ }\textbf {\bibinfo {volume} {94}},\ \bibinfo {pages}
  {076406} (\bibinfo {year} {2005})}\BibitemShut {NoStop}%
\bibitem [{\citenamefont {Ivchenko}(1973)}]{Ivchenko1973}%
  \BibitemOpen
  \bibfield  {author} {\bibinfo {author} {\bibfnamefont {E.~L.}\ \bibnamefont
  {Ivchenko}},\ }\bibfield  {title} {\enquote {\bibinfo {title} {{Spin relaxation of free carriers in
  semiconductors without inversion center in longitudinal magnetic field}},}\
  }\href@noop {} {\bibfield  {journal} {\bibinfo  {journal} {Fiz. Tverd. Tela}\
  }\textbf {\bibinfo {volume} {15}},\ \bibinfo {pages} {1566} (\bibinfo {year}
  {1973}) [Sov. Phys.
  Solid State 15, 1048 (1973)]}\BibitemShut {NoStop}%
\bibitem [{\citenamefont {Glazov}\ and\ \citenamefont
  {Ivchenko}(2002)}]{Glazov02}%
  \BibitemOpen
  \bibfield  {author} {\bibinfo {author} {\bibfnamefont {M.~M.}\ \bibnamefont
  {Glazov}}\ and\ \bibinfo {author} {\bibfnamefont {E.~L.}\ \bibnamefont
  {Ivchenko}},\ }\bibfield  {title} {\enquote {\bibinfo {title} {{Precession
  spin relaxation mechanism caused by frequent electron-electron
  collisions}},}\ }\href {\doibase 10.1134/1.1490009} {\bibfield  {journal}
  {\bibinfo  {journal} {JETP Lett.}\ }\textbf {\bibinfo {volume} {75}},\
  \bibinfo {pages} {403} (\bibinfo {year} {2002})}\BibitemShut {NoStop}%
\bibitem [{\citenamefont {Glazov}\ and\ \citenamefont
  {Ivchenko}(2004)}]{Glazov04}%
  \BibitemOpen
  \bibfield  {author} {\bibinfo {author} {\bibfnamefont {M.~M.}\ \bibnamefont
  {Glazov}}\ and\ \bibinfo {author} {\bibfnamefont {E.~L.}\ \bibnamefont
  {Ivchenko}},\ }\bibfield  {title} {\enquote {\bibinfo {title} {{Effect of
  electron-electron interaction on spin relaxation of charge carriers in
  semiconductors}},}\ }\href {\doibase 10.1134/1.1854815} {\bibfield  {journal}
  {\bibinfo  {journal} {J. Exp. Theor. Phys.}\ }\textbf {\bibinfo {volume}
  {99}},\ \bibinfo {pages} {1279} (\bibinfo {year} {2004})}\BibitemShut
  {NoStop}%
\bibitem [{\citenamefont {Marushchak}\ \emph {et~al.}(1983)\citenamefont
  {Marushchak}, \citenamefont {Stepanova},\ and\ \citenamefont
  {Titkov}}]{Marushchak1983}%
  \BibitemOpen
  \bibfield  {author} {\bibinfo {author} {\bibfnamefont {V.~A.}\ \bibnamefont
  {Marushchak}}, \bibinfo {author} {\bibfnamefont {M.~N.}\ \bibnamefont
  {Stepanova}}, \ and\ \bibinfo {author} {\bibfnamefont {A.~N.}\ \bibnamefont
  {Titkov}},\ }\bibfield  {title} {\enquote {\bibinfo {title} {{Suppression by
  a longitudinal magnetic field of spin relaxation of conduction electrons in
  semiconductor crystals lacking an inversion center}},}\ }\href@noop {}
  {\bibfield  {journal} {\bibinfo  {journal} {JETP Lett.}\ }\textbf {\bibinfo
  {volume} {37}},\ \bibinfo {pages} {400} (\bibinfo {year} {1983})}\BibitemShut
  {NoStop}%
\bibitem [{\citenamefont {Jusserand}\ \emph {et~al.}(1995)\citenamefont
  {Jusserand}, \citenamefont {Richards}, \citenamefont {Allan}, \citenamefont
  {Priester},\ and\ \citenamefont {Etienne}}]{Jusserand1995}%
  \BibitemOpen
  \bibfield  {author} {\bibinfo {author} {\bibfnamefont {B.}\ \bibnamefont
  {Jusserand}}, \bibinfo {author} {\bibfnamefont {D.}\ \bibnamefont
  {Richards}}, \bibinfo {author} {\bibfnamefont {G.}\ \bibnamefont {Allan}},
  \bibinfo {author} {\bibfnamefont {C.}\ \bibnamefont {Priester}}, \
  and\ \bibinfo {author} {\bibfnamefont {B.}\ \bibnamefont {Etienne}},\
  }\bibfield  {title} {\enquote {\bibinfo {title} {{Spin orientation at
  semiconductor heterointerfaces}},}\ }\href {\doibase
  10.1103/PhysRevB.51.4707} {\bibfield  {journal} {\bibinfo  {journal} {Phys.
  Rev. B}\ }\textbf {\bibinfo {volume} {51}},\ \bibinfo {pages} {4707}
  (\bibinfo {year} {1995})}\BibitemShut {NoStop}%
\bibitem [{\citenamefont {Richards}\ and\ \citenamefont
  {Jusserand}(1999)}]{Richards1999}%
  \BibitemOpen
  \bibfield  {author} {\bibinfo {author} {\bibfnamefont {D.}~\bibnamefont
  {Richards}}\ and\ \bibinfo {author} {\bibfnamefont {B.}~\bibnamefont
  {Jusserand}},\ }\bibfield  {title} {\enquote {\bibinfo {title} {{Spin
  energetics in a GaAs quantum well: Asymmetric spin-flip Raman scattering}},}\
  }\href {\doibase 10.1103/PhysRevB.59.R2506} {\bibfield  {journal} {\bibinfo
  {journal} {Phys. Rev. B}\ }\textbf {\bibinfo {volume} {59}},\ \bibinfo
  {pages} {R2506} (\bibinfo {year} {1999})}\BibitemShut {NoStop}%
\bibitem [{\citenamefont {Bronold}\ \emph {et~al.}(2002)\citenamefont
  {Bronold}, \citenamefont {Martin}, \citenamefont {Saxena},\ and\
  \citenamefont {Smith}}]{Bronold2002}%
  \BibitemOpen
  \bibfield  {author} {\bibinfo {author} {\bibfnamefont {F.~X.}\
  \bibnamefont {Bronold}}, \bibinfo {author} {\bibfnamefont {I.}\
  \bibnamefont {Martin}}, \bibinfo {author} {\bibfnamefont {A.}\
  \bibnamefont {Saxena}}, \ and\ \bibinfo {author} {\bibfnamefont {D.~L.}\
  \bibnamefont {Smith}},\ }\bibfield  {title} {\enquote {\bibinfo {title}
  {{Magnetic-field dependence of electron spin relaxation in $n$-type
  semiconductors}},}\ }\href {\doibase 10.1103/PhysRevB.66.233206} {\bibfield
  {journal} {\bibinfo  {journal} {Phys. Rev. B}\ }\textbf {\bibinfo {volume}
  {66}},\ \bibinfo {pages} {233206} (\bibinfo {year} {2002})}\BibitemShut
  {NoStop}%
\bibitem [{\citenamefont {Gor'kov}\ \emph {et~al.}(1979)\citenamefont
  {Gor'kov}, \citenamefont {Larkin},\ and\ \citenamefont
  {Khmel'nitskii}}]{Gorkov1979}%
  \BibitemOpen
  \bibfield  {author} {\bibinfo {author} {\bibfnamefont {L.~P.}\ \bibnamefont
  {Gor'kov}}, \bibinfo {author} {\bibfnamefont {A.~I.}\ \bibnamefont {Larkin}},
  \ and\ \bibinfo {author} {\bibfnamefont {D.~E.}\ \bibnamefont
  {Khmel'nitskii}},\ }\bibfield  {title} {\enquote {\bibinfo {title}
  {{Particle conductivity in a two-dimensional random potential}},}\ }\href
  {\doibase 10.1142/9789814317344\_0022} {\bibfield  {journal} {\bibinfo
  {journal} {JETP Lett.}\ }\textbf {\bibinfo {volume} {30}},\ \bibinfo {pages}
  {228} (\bibinfo {year} {1979})}\BibitemShut {NoStop}%
\bibitem [{\citenamefont {Dmitriev}\ \emph {et~al.}(1997)\citenamefont
  {Dmitriev}, \citenamefont {Kachorovskii},\ and\ \citenamefont
  {Gornyi}}]{Dmitriev1997}%
  \BibitemOpen
  \bibfield  {author} {\bibinfo {author} {\bibfnamefont {A.~P.}\ \bibnamefont
  {Dmitriev}}, \bibinfo {author} {\bibfnamefont {V.~Yu.}\ \bibnamefont
  {Kachorovskii}}, \ and\ \bibinfo {author} {\bibfnamefont {I.~V.}\
  \bibnamefont {Gornyi}},\ }\bibfield  {title} {\enquote {\bibinfo {title}
  {{Nonbackscattering contribution to weak localization}},}\ }\href {\doibase
  10.1103/PhysRevB.56.9910} {\bibfield  {journal} {\bibinfo  {journal} {Phys.
  Rev. B}\ }\textbf {\bibinfo {volume} {56}},\ \bibinfo {pages} {9910}
  (\bibinfo {year} {1997})}\BibitemShut {NoStop}%
\bibitem [{\citenamefont {Emel'yanenko}\ \emph {et~al.}(1982)\citenamefont
  {Emel'yanenko}, \citenamefont {Lagunova},\ and\ \citenamefont
  {Polyanskaya}}]{Emelyanenko1982}%
  \BibitemOpen
  \bibfield  {author} {\bibinfo {author} {\bibfnamefont {O.V.}\ \bibnamefont
  {Emel'yanenko}}, \bibinfo {author} {\bibfnamefont {T.S.}\ \bibnamefont
  {Lagunova}}, \ and\ \bibinfo {author} {\bibfnamefont {T.A.}\ \bibnamefont
  {Polyanskaya}},\ }\bibfield  {title} {\enquote {\bibinfo {title} {{Mechanism
  for relaxation of the phase of the electron wave function in \$n\$-type
  GaAs}},}\ }\href@noop {} {\bibfield  {journal} {\bibinfo  {journal} {JETP
  Lett.}\ }\textbf {\bibinfo {volume} {36}},\ \bibinfo {pages} {246} (\bibinfo
  {year} {1982})}\BibitemShut {NoStop}%
\bibitem [{\citenamefont {Isawa}(1984)}]{Isawa1984}%
  \BibitemOpen
  \bibfield  {author} {\bibinfo {author} {\bibfnamefont {Y.}\
  \bibnamefont {Isawa}},\ }\bibfield  {title} {\enquote {\bibinfo {title}
  {{Inelastic Scattering Time in Disordered Metals}},}\ }\href {\doibase
  10.1143/JPSJ.53.2865} {\bibfield  {journal} {\bibinfo  {journal} {J. Phys.
  Soc. Japan}\ }\textbf {\bibinfo {volume} {53}},\ \bibinfo {pages}
  {2865} (\bibinfo {year} {1984})}\BibitemShut {NoStop}%
\bibitem{Suppl} See Supplemental Material at \href {http://link.aps.org/supplemental/10.1103/PhysRevX.8.031021}{http://link.aps.org/supplemental/10.1103/PhysRevX.8.031021} for more details on the calculation of the relation between $\tau_p$ and $\tau_3$ and on the experimental results related to the high-density sample.
\end{thebibliography}
\end{document}